 \documentclass[final,3p,times]{elsarticle}

\usepackage{amssymb}
\usepackage{amsmath}
\usepackage{mdwmath}
\usepackage{mdwtab}

\usepackage{verbatim}
\usepackage[caption=false,font=footnotesize]{subfig}

\journal{}

\begin{document}

\begin{frontmatter}




 \title{Closed-form, robust and accurate multi-frequency phase unwrapping: frequency design and algorithm\tnoteref{label3}}

  \tnotetext[label3]{This work has been supported by the National Natural Science Foundation
 of China (No.61402520; No.61273047; No.61573376  ) and the Natural Science Foundation of Jiangsu Province (BK20130068).}


    \author[]{Li~Wei\corref{}  }
    \ead{wlnb@hotmail.com}

    \author[]{~Wangdong~Qi\corref{cor1}  }%
    \ead{wangdongqi@gmail.com}


   \cortext[cor1]{Corresponding author}

 \address{PLA University of Science and Technology, Nanjing, China}






\begin{abstract}

     A closed-form algorithm, named ``concerto'', is proposed for phase-based distance estimation in multi-frequency phase unwrapping (MFPU) system. The \textit{concerto} method consists of three coherent estimation stages,i.e., initial modified BW estimation, residual error estimation and LS estimation ,
     each of which has a closed-form expression and cooperates closely with each other like a concerto. Due to a specially designed frequency pattern, \textit{concerto} is reliable, accurate, and computationally simple.
     Meanwhile, measurement frequency selection is an easier task.
     Performance comparisons with beat wavelength (BW), excess fractions (EF) and Chinese remainder theorem (CRT) method confirm that our method outperforms these methods both in accuracy and reliability and can asymptotically achieve the Cram\'{e}r-Rao bound (CRB).

\end{abstract}


\end{frontmatter}


\section{Introduction}
\label{}

   Precise distance or height measurement is of great importance for many fields such as geodesy\cite{Kusy:06,Wang:11}, synthetic aperture radar (SAR)\cite{Xia:07,Li:08} or interferometric synthetic aperture radar (InSAR) \cite{Yuan:13} and optics\cite{Towers:03,Falaggis:11}. In these cases, phase measurements at multiple frequencies are used for accurate estimation of distance or height. The obstacle met in MFPU is that the measured phases are wrapped into the range $(-\pi,\pi]$, while the true distance is related to the unwrapped phases. To recover the unwrapped phases, unknown integer called folding integer must be determined by a phase unwrapping method.

    A least-square grid search is used to solve the phase unwrapping problem in \cite{Kusy:06}. Since the processing time is dependent on both the range and the search step, it is usually computationally prohibitive for real-time applications. If the measurement wavelengths can be scaled to integers and these integers are pairwise co-prime, the CRT algorithm may be applied\cite{Xia:07,Li:08,Wang:10,Wang:11}. Although the traditional CRT is computationally attractive owing to its closed-form solution, it is very sensitive to phase noise\cite{Huang:87}. More importantly, the frequency selection is a challenge since the co-prime condition must be met for any pair of frequencies \cite{Wang:11}. In order to solve the noise sensitivity problem, a two-dimensional searching based robust CRT has been proposed in \cite{Xia:07}. Thus, there exists a compromise between reliability and complexity. The two-dimensional searching used in  \cite{Xia:07} is later reduced to one-dimensional searching by \cite{Li:08}. To further reduce the computational complexity of searching based CRT, closed-form and robust CRT are presented in \cite{Wang:10,Yuan:13}. Recently, new closed-form phase unwrapping using lattice theory is proposed in \cite{Li:13}, which has similar complexity and accuracy as the CRT in \cite{Wang:10}.
    But the rigid requirement on measurement frequencies still exists \cite{Wang:10,Li:13}. The robust CRT proposed in \cite{Wang:11} both alleviates the requirement on frequencies and improves the resistance to noise,  at the cost of very limited measurement range.
    The classical EF method has also been proposed for decades, see \cite{Falaggis:11} and references therein, which searches all the possible locations determined at the shortest wavelength. At the most likely location, all folding integers calculated are closest to integer simultaneously\cite{Falaggis:11}. Because the processing time is increased linearly with range, EF approach is still computationally intensive. Compared with EF, the popular BW method has similar reliability, yet requires very low computation cost\cite{Towers:03}.\par

    However, BW estimates the folding integer one by one and leads to accuracy loss since only partial information is used for each folding integer estimation. The frequency pattern designed for BW may cause poor estimation performance in case of large number of frequencies\cite{Towers:03,wei:12a}.
    In this letter, another \textit{closed-form} phase unwrapping method, named \textit{concerto}, is provided.

    The main features of this algorithm include:
    \begin{enumerate}
      \item[i)]  Large (Adjustable) measurable range, low computation complexity, high reliability and accuracy can be achieved simultaneously.
      \item[ii)] Frequency selection is easier than CRT based method.
    \end{enumerate}

\section{Signal model and BW method}
\label{sect:model}
 Consider a  multi-frequency ranging system using phase measurements recorded at multiple wavelengths \cite{Wang:11,Li:13}.
   Assume the wrapped phase (or principal phase) measurements at $N$ wavelengths $\lambda_0<\lambda_1\cdots <\lambda_{N-1}$ are $\phi_0\cdots \phi_{N-1}$.  In MFPU, the ideal noise-free measurement phases are related to the range $L$ by
   \footnote{Note that similar problems and signal model arise in optic \cite{Towers:03}, SAR \cite{Li:08}, InSAR imaging system \cite{Yuan:13}, frequency estimation \cite{Li:15} and single source direction of arrival (DOA) estimation\cite{wei:15}.  }
   \begin{gather}
    \label{equa:basic}
        \phi_i=\left[2\pi\frac{L}{\lambda_i} \right]_{2\pi},\;0\leq i\leq N-1
    \end{gather}

   In BW method, using the general formula
   \begin{gather*}
       \left[[x]_{2\pi}-[y]_{2\pi}\right]_{2\pi}=[x-y]_{2\pi}
   \end{gather*}
   the beat (synthetic) phases $\Phi_{i}$ and beat wavelengths $\Lambda_{i}$ can be formed as
   \begin{align}
   \label{equa:beat_phase}
       \Phi_{i}&=[\phi_0-\phi_i]_{2\pi}=\left[2\pi\frac{L}{\lambda_0}-2\pi\frac{L}{\lambda_i}\right]_{2\pi}
       =\left[2\pi\frac{L}\Lambda_{i}\right]_{2\pi}\\
       \Lambda_{i}&=\lambda_i\lambda_0/(\lambda_i-\lambda_0)=c/(f_0-f_i) ,\quad 1\leq i\leq N-1
   \end{align}
   where $\lambda_i=c/f_i$ and $c$ is the speed of light, $[\cdot]_{2\pi}$ denotes a {modulo-$2\pi$} operation.
   Equation (\ref{equa:basic}) is equivalent to
   \begin{align}\label{equa:L_m}
     L&=\left(m_i+{\phi_i}/{2\pi}\right)\lambda_i,\quad 0\leq i\leq N-1
   \end{align}
   and (\ref{equa:beat_phase}) can be expressed as
   \begin{align}\label{equa:L_M}
     L&=\left(M_i+{\Phi_{i}}/{2\pi}\right)\Lambda_{i}\notag\\
      &=\left(M_{i+1}+{\Phi_{i+1}}/{2\pi}\right)\Lambda_{i+1},\quad 1\leq i\leq N-2
   \end{align}
   where $m_i$, $M_i$ are the folding integers at wavelength $\lambda_i$ and beat wavelength $\Lambda_{i}$ respectively.
   Note that the unambiguous measurement range (UMR) is equal to the largest beat wavelength $\Lambda_{1}$ in BW method and the unknown range $L$ will satisfy $|L|<\Lambda_{1}/2=\frac{c}{2(f_0-f_1)}$ by proper choice of $\Lambda_{1}$.
   The same assumption is made in this letter.
   Then we have $M_1=0$ immediately. Now, consider the realistic measurement phases with phase noise $\theta_e (i)$,
   \begin{gather}
   \label{equa:basic_noise}
       \phi_i=\left[2\pi\frac{L}{\lambda_i} +\theta_e (i) \right]_{2\pi},\; 0\leq i\leq N-1
   \end{gather}
   using (\ref{equa:L_m}) and (\ref{equa:L_M}), the integers $M_i$, $m_0$ can be calculated sequentially as follows:
   \begin{align}%
    \label{equa:M_estimate}
    \hspace{-8pt}M_{i+1}&=\textrm{round}\left[\left( M_i+\frac{\Phi_{i}}{2\pi}\right)\frac{\Lambda_{i}}{\Lambda_{i+1}}-\frac{\Phi_{i+1}}{2\pi} \right]   \\
    \label{equa:m_estimate}
    m_0&=\textrm{round}\left[\left( M_{N-1}+\frac{\Phi_{N-1}}{2\pi}\right)\frac{\Lambda_{N-1}}{\lambda_0}-\frac{\phi_0}{2\pi} \right]
   \end{align}
   where $\mathrm{round}[\cdot]$ denotes rounding to the nearest integer. It follows that
   \begin{align}
     L=\left(m_0+{\phi_0}/{2\pi}\right)\lambda_0
   \end{align}

\section{Frequency design}

    Suppose the phase noise $\theta_e (i)$ at each wavelength $\lambda_i$ is zero-mean Gaussian noise with identical standard deviation $\sigma$, then the error in (\ref{equa:M_estimate}), before the rounding operation, is also zero-mean Gaussian noise with deviation $\sigma_e$
    \footnote{There may be other phase error, i.e. due to multipath, which we could cancel by antenna design or multi-frequency average\cite{zhang:14}, this is out of the scope of the paper.},
   \begin{gather}\label{equa:noise_amplify}
      \sigma_e=\frac{\sqrt{2}\sigma}{2\pi} \sqrt{\left(\frac{\Lambda_{i}}{\Lambda_{i+1}}\right)^2+1}
   \end{gather}
   According to \cite{Towers:03}, to maximize noise immunity, the phase noise introduced in each $M_i$ estimation of (\ref{equa:M_estimate}) should have identical standard deviation, and thus the scaling factor ${\Lambda_{i}}/{\Lambda_{i+1}}$ must be all equal to each other
   \footnote{This principle can be interpreted using the "Barrel Theory", which states that the capacity of a barrel is limited by the shortest stave.
   Therefore, if the $M_i$ estimation has been corrupted by the maximal phase noise, then it is "the shortest stave" and is prone to error.
   When the error happens, it propagates along the estimation chain in (\ref{equa:M_estimate}).}.

   Therefore, the frequency pattern designed for BW method is \cite{Towers:03}
   \begin{align}
      \frac{\Lambda_{1}}{\Lambda_{2}}&=\frac{\Lambda_{2}}{\Lambda_{3}}=\cdots =\frac{\Lambda_{N-1}}{\lambda_0}
   \end{align}
   or equivalently,
   \begin{align}
    \label{equa:scale_equal}
    \frac{f_0-f_2}{f_0-f_1}&=\frac{f_0-f_3}{f_0-f_2}=\cdots =\frac{f_0}{B}
   \end{align}
   where $B=f_0-f_{N-1}$ denotes the measurement bandwidth.

   Note that the last equality in (\ref{equa:scale_equal}) is needed to unwrap $m_0$ using (\ref{equa:m_estimate}) since it is the final step in the estimation chain.
   The final absolute error will be smaller than $\lambda_0/2$ when $m_0$ has been correctly unwrapped. However, the last step of calculating $m_0$ in (\ref{equa:m_estimate}) is \textit{not essential}  if the method to be described in the next section is used. Then, the constraint of ${\Lambda_{i}}/{\Lambda_{i+1}}=\frac{f_0}{B}$ could be removed. This is a key step to achieve high-accuracy estimation.
   Since $\frac{f_0}{B}$ is usually large, the measurement frequencies will converge quickly to $f_0$ with increased frequency number.
   This kind of frequency pattern will lead to poor estimation accuracy, see \cite{wei:12a} for detail. The proposed frequency pattern used in \textit{concerto}, takes the form
    \begin{align}
    \label{equa:scale_new}
      \hspace{-4pt}r=\frac{f_0-f_2}{f_0-f_1}=\frac{f_0-f_3}{f_0-f_2}=\cdots =\frac{f_0-f_{N-1}}{f_0-f_{N-2}}
    \end{align}
    multiplying the last $N-1-i$ terms of (\ref{equa:scale_new})
    \begin{align}
      \hspace{-4pt}f_i=f_0-(f_0-f_{N-1})r^{-(N-1-i)},\; i=1,2,\cdots,N-2
    \end{align}
    where $f_0$ and $f_{N-1}$ are the pre-determined highest and lowest frequency.
    Suppose $|L|<K/2$, $K$ is the maximum measurable range of ranging system. Since we require $UMR\geq K$ with $UMR=\Lambda_{1}=\frac{c}{f_0-f_1}$,
    the ratio $r$ is adjusted adaptively according to the frequency number $N$,
    \begin{gather}%
      \label{}
      r^{N-2}=\frac{f_0-f_{N-1}}{f_0-f_1}=\frac{B\cdot UMR}{c}\geq \frac{B\cdot K}{c} \notag\\
       r\geq \sqrt[N-2]{\frac{B\cdot K}{c}}
    \end{gather}
    We set $r=\sqrt[N-2]{{B\cdot K}/{c}}$ for maximizing the tolerance of noise, see (\ref{equa:noise_amplify}). Frequency pattern design is then easily accomplished  by simply setting $r$ in \textit{concerto}.

   \textit{Remark 1}: Under the constraint of being larger than unity, the ratios in (\ref{equa:scale_new}) are required to be not only equal, but also as small as possible. Since $UMR=\frac{B}{c}r^{N-2}$, even a ratio slightly greater than unity will produce an extremely large $UMR$ thanks to the exponent increase property. Error accumulation is avoided by decision at each step of (\ref{equa:M_estimate}).


\section{Proposed Method}


   In this section, we present the \textit{concerto} method which adds two optimal estimations to a modified BW method (using the proposed frequency pattern) and achieves high accuracy with extreme low computation cost due to closed-form solution developed. The modified BW method still suffers from accuracy loss since only partial information is exploited in each phase unwrapping stage using just two synthetic wavelengths, see (\ref{equa:M_estimate}).
   With two additional steps, \textit{concerto} aims to recover the \textit{lost  information}  by making full use of all the phase
   information.

   We use (\ref{equa:M_estimate}) for a rough estimation of $L$ in the first of three stages.
   The estimation error of the last step in (\ref{equa:M_estimate}) lies in $\left[-c/2(f_0-f_{N-1}),c/2(f_0-f_{N-1})\right]=\left[-c/(2B),c/(2B)\right]$, provided that $M_{N-1}$ has been correctly estimated.
   Then a coarse estimate $L_c$ is
   \begin{gather}
      \label{equa:L_ini}
       {L_c}=M_{N-1}\Lambda_{N-1}+\frac{\Phi_{N-1}}{2\pi}\Lambda_{N-1}
   \end{gather}
   The original wrapped phase $\phi_i$ will be compensated as follows
   \begin{align}\label{}
       \tilde{\phi}_{i}=\left[\phi_i-\frac{2\pi{L_c}}{\lambda_i}\right]_{2\pi}
                =\left[\frac{2\pi(L-L_c)}{\lambda_i}+ \theta_e (i)\right]_{2\pi}
   \end{align}

   Denote the residual error $L_r=L-L_c$.
   To obtain $L_r$, we construct the following cost function
      \begin{align}
        \label{equa:cost_max}
         &\underset{\hat{L}_r}{\max} \left|\sum_{i=0}^{N-1} \hspace{-2pt}\exp \hspace{-2pt}\left \{{j\left(\frac{2\pi \hat{L}_r}{\lambda_i}- \tilde{\phi}_{i} \right)} \right \}    \right|^2  \notag\\
         &=\underset{\hat{L}_r}{\max} \left|\sum_{i=0}^{N-1} \hspace{-2pt}\exp \hspace{-2pt}\left \{ j\left(\frac{2\pi f_i \hat{L}_r}{c}-\tilde{\phi}_{i} \right) \right \}    \right|^2
      \end{align}
   where $\hat{L}_r$ is the estimation of $L_r$.
   This is because when $\hat{L}_r=L_r$, all the unit-vector $e^{ j\left({2\pi f_i \hat{L}_r}/{c}-\tilde{\phi}_{i} \right) }$, $i=0,\cdots,N-1$, will be aligned to the same direction and the cost function will achieve its maximum value.


   Let ${L_r^*}$ be the optimal estimation of $L_r$.
   When $|L_r|<c/2B$
   \footnote{ Eq.(\ref{equa:L_wpa}) does not hold true for $|L_r|>c/2B$, see Appendix \ref{append:optimal_estimation} for details. Therefore, $|L_r|<c/2B$, guaranteed by the first stage of \textit{concerto}, forms the core of the second stage.},
   ${L_r^*}$ must be of the form (see \ref{append:optimal_estimation})
   \begin{align}
    \label{equa:L_wpa}
     {L_r^*} &= \frac{c}{2\pi} \frac{\Delta \mathbf{f}^T \mathbf{\Gamma}^T \mathbf{R}^{-1} \mathbf{\Gamma}\Delta\tilde{\mathbf{\Phi}} }
                  { \Delta \mathbf{f}^T \mathbf{\Gamma}^T \mathbf{R}^{-1} \mathbf{\Gamma}\Delta \mathbf{f}  }
   \end{align}
   where
   $\Delta \mathbf{f}=[ \Delta f_1 ,\Delta f_2 , \cdots\Delta f_{N-1} ]^T$, $\Delta f_i=f_{i-1}-f_{i}$,
   $\Delta\tilde{\mathbf{\Phi}}=[\Delta\tilde{\phi}_{0,1},\Delta\tilde{\phi}_{1,2},\cdots,\Delta\tilde{\phi}_{N-2,N-1}]$,
   $\Delta\tilde{\phi}_{i,k}=\left[\tilde{\phi}_{i}-\tilde{\phi}_{k}\right]_{2\pi}$,
   and $\mathbf{R}^{-1}$, $\mathbf{\Gamma}$ are defined as
   \begin{gather}
     \mathbf{\Gamma}= \left[\begin{array}{*{20}c}
                       1 &    0     & \cdots   & 0  \notag\\
                       1 &    1     & \cdots   & 0  \notag\\
                 \vdots  &  \vdots  &  1       & 0  \notag\\
                       1 &    1     &  1       & 1  \notag\\
                     \end{array}  \right]           \notag\\
      \label{equa:R_u}
      \mathbf{R}^{-1}=\mathbf{I}_{N-1}-\frac{\mathbf{uu}^T}{N},\,
      \mathbf{u}=[1,1,\cdots 1]^T
   \end{gather}
   where $\mathbf{I}_N$ denotes a $N\times N$ identity matrix.
   Let $\mathbf{W}=\mathbf{\Gamma}^T \mathbf{R}^{-1} \mathbf{\Gamma}$, which is a constant, with the $(j,k)$ entry
   $\mathbf{W}_{jk}$,
   \begin{gather*}
       \mathbf{W}_{jk}=\frac{N{\rm min}(j,k)-jk}{N}, 1\leq j,k\leq N-1
   \end{gather*}
   Then, (\ref{equa:L_wpa}) can be simplified as
   \begin{align}
    \label{equa:L_wpa_simp}
     {L_r^*} &= \frac{c}{2\pi} \frac{\Delta \mathbf{f}^T \mathbf{W} \Delta\tilde{\mathbf{\Phi}} }
                  { \Delta \mathbf{f}^T \mathbf{W} \Delta \mathbf{f}  }
   \end{align}
    The optimal estimation of the second stage is readily obtained
   \begin{gather}\label{equa:coarse}
     L_m=L_r^*+L_c
   \end{gather}
   If the residual error $|L-L_m|$ of (\ref{equa:coarse}) is small enough, and the folding integer vector
   $\mathbf{\hat{m}}=[\hat{m}_0,\hat{m}_1,\cdots,\hat{m}_{N-1}]$ could be estimated correctly by
    \begin{gather}\label{}
        \hat{m}_i=\textrm{round}\left[\frac{L_m}{\lambda_i}-\frac{\phi_i}{2\pi}\right]
    \end{gather}
   Then the least square (LS) based cost function in MFPU could be written as
   \begin{align}\label{}
      J(L)&=\sum_{i=0}^{N-1}\left( \left[\frac{2\pi L}{\lambda_i}-\phi_i\right]_{2\pi}  \right)^{2} \notag\\
                       &=\sum_{i=0}^{N-1}\left(    \frac{2\pi L}{\lambda_i}-\phi_i-2\pi \hat{m}_i \right)^{2}
   \end{align}
   Denote $\Lambda_{inv}=[\frac{1}{\lambda_0},\frac{1}{\lambda_1},\cdots,\frac{1}{\lambda_{N-1}}]$ and $\mathbf{m_f}=[m_{f0},m_{f1},\cdots,m_{fN-1}]$, $m_{fi}=\hat{m}_i+\frac{\phi_i}{2\pi}$, then
   \begin{align}\label{}
       J(L)=4\pi^2\left( \Lambda_{inv}L-\mathbf{m_f} \right)\left( \Lambda_{inv}L-\mathbf{m_f} \right)^T
   \end{align}
   The optimal estimation $L^*$ of \textit{concerto} that minimizing $J(L)$ is finally obtained
    \begin{align}\label{equa:opt_estimation}
        L^*=\frac{\Lambda_{inv}\mathbf{m_f}^T}{\Lambda_{inv}\Lambda_{inv}^T}
    \end{align}
   It is easy to verify that
    \begin{align}
        &\Delta L =L^*-L=\frac{\sum_{k=0}^{N-1} \lambda_k^{-1} \theta_e (k) } {2\pi \sum_{k=0}^{N-1} {\lambda_k^{-2}} }\\
        &\mathbf{E} \left[\Delta L \right]=0 \\
        \label{equa:opt_MSE}
        &\mathbf{E} \left[(\Delta L)^2 \right]=\frac{1}{4\pi^2}\frac{\mathbf{E} \left[ \left(\sum_{k=0}^{N-1} \lambda_k^{-1} \theta_e (k) \right)^2 \right]}{ \left(\sum_{k=0}^{N-1} {\lambda_k^{-2}}\right)^2 }\notag\\
        &\hspace{45pt}=\frac{\sigma ^2}{4\pi^2 \sum_{k=0}^{N-1} {\lambda_k^{-2}} }
    \end{align}
   where $\mathbf{E}[\cdot]$ is the expectation operation and $\mathbf{E}[\theta_e (k)\theta_e (j)]=0$ for $k\neq j$.

\textit{Remark 2}: Note that no matrix inversion or matrix decomposition is required in the computation of $L^*$. So, \textit{concerto} is suitable for hardware implementation.



   The Cramer-Rao bound (CRB) of $L$ is also shown for comparison (see \ref{append:CRB})
     \begin{align}
     \label{equa:CRB}
        \mathrm{CRB}(L) = \frac{\sigma ^2}{4\pi^2}\left(\sum_{k=1}^N {\lambda_k^{-2}} \right)^{-1}
     \end{align}

   From Eqs.(\ref{equa:opt_MSE}) and (\ref{equa:CRB}), it is worth noting that the estimator given by Eqs. (\ref{equa:L_ini})-(\ref{equa:opt_estimation}) has attained the optimal accuracy by utilizing all the phase information simultaneously.

\section{Simulation Analysis}

   Unless otherwise mentioned, simulations are performed under the following conditions:
   For fair comparison of the robust dual-frequency CRT (DCRT) \cite{Wang:11}, EF, BW and \textit{concerto} method, simulations are compared under the same bandwidth and frequency number. The highest and lowest frequency are set as $f_0=2500\textrm{\,MHz}$, $f_{N-1}=2400\textrm{\,MHz}$, with $B=100\textrm{\,MHz}$, and $N=51$. For DCRT method, the frequency pattern is designed as in \cite{Wang:11} with quantization step $u=0.0001 \mathrm{\,m}$, $C=1$ and $R=30 \mathrm{\,m}$. For the other methods, the proposed frequency pattern is used with $K=144 \mathrm{\,m}$ (the same as that in \cite{Wang:11}) and $r=\sqrt[49]{{B\cdot K}/{c}}$. SNR is defined as $\textrm{SNR} =\frac{1}{2\mathbf{E} [\theta_e^2 (k)]}=\frac{1}{2 \sigma^2 }$.

  \begin{table}[tb]
    \renewcommand{\arraystretch}{1.3}  
    \caption{Computation cost of the EF, BW, DCRT and concerto method using the frequency pattern
    described below (both $K=144\mathrm{\,m}$ and $K=14400\mathrm{\,m}$ are evaluated )}
    \label{table_compute_cost}
    \centering      
    \begin{tabular}{|l|c|c|c|c|}\hline  
                 & Range        & Process Time           & Range         & Process Time  \\
          Method &({\textrm{m}})&   (ms)               & ({\textrm{m}})&   (ms)        \\\hline
          EF     &  144         & 7.62                   & 14400         & 862.15        \\\hline
          BW     &  144         & 0.11                   & 14400         & 0.12          \\\hline
          DCRT    &  144         & 3.59                   & 14400         & N/A           \\\hline
          \textit{concerto}    &  144         & 0.21                   & 14400         & 0.23          \\\hline
   \end{tabular}
   \end{table}

    In Table \ref{table_compute_cost}, the relative computational load is represented by process time collected in Matlab, running on a 2.33GHz processor with 2GB memory. The BW and \textit{concerto} method have comparable complexity, which both are almost independent of range and lower than DCRT or EF method. As expected, the computation cost of EF is much higher than the others and linearly increased with range. Note that the process time of DCRT is not simulated for larger range because of its limited measurement range.

\begin{figure}[t]  
       \centering{\includegraphics[width=85mm,height=65mm]{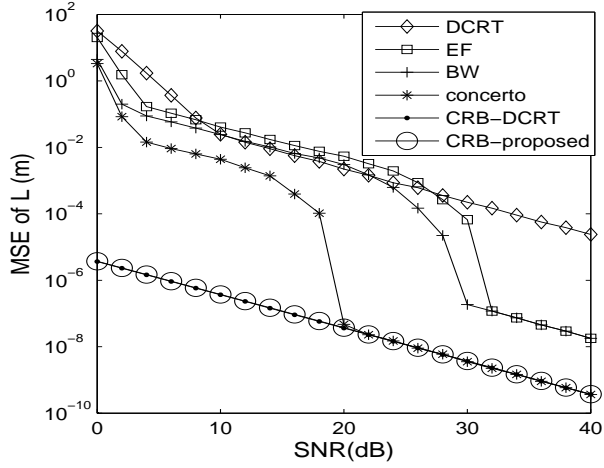}}
       \caption{ MSE versus SNR for different methods with $N=51$ and $B=100\textrm{\,MHz}$.
         }
       \label{fig_MSE}
       \vspace{-0pt}
    \end{figure}

    The mean square error (MSE) of different methods, defined as $MSE(L)=\mathbf{E}[(\hat{L}-L)^2]=\mathbf{E}[(\Delta L)^2]$, where $\hat{L}$ is the estimated value of the true $L$,
    are shown in Fig.\ref{fig_MSE}. It reveals that \textit{concerto} remarkably outperforms the others in MSE performance for the whole SNR region simulated.
    The CRB for the frequency pattern in \cite{Wang:11} and the proposed frequency pattern are also shown (denoted as CRB-DCRT and CRB-proposed respectively).
    Both CRB are observed to be almost completely overlapped.
    Furthermore, only the proposed approach asymptotically attains the CRB for high SNR, thus possessing the optimal accuracy and validating the theoretical analysis in (\ref{equa:opt_MSE}).

\begin{figure}[t]   
         \centering{\includegraphics[width=85mm,height=65mm]{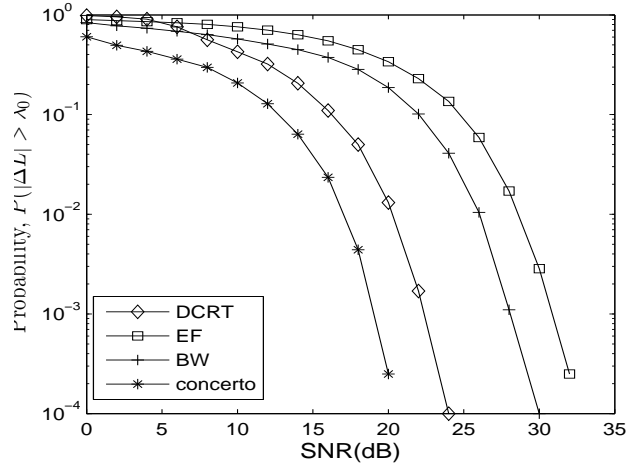}}
         \caption{Probability of the absolute error larger than the wavelength $\lambda_0$ with $N=51$ and $B=100\textrm{\,MHz}$.
           }
      \label{fig_Pf}
      \vspace{-0pt}
    \end{figure}

    The probability $P(|\Delta L|>\lambda_0)$ is of particular interest for accurate ranging. In Fig.\ref{fig_Pf}, the reliability of \textit{concerto}, represented by the probability $P(|\Delta L|>\lambda_0)$, is far more superior to the others.
    For example, compared with the others, about 3 dB to even 10 dB gain is obtained for \textit{concerto} under the same reliability.
    It is interesting to note that the robust DCRT method has poor MSE performance but with relatively better reliability.

    \begin{figure}[!t]
     \centerline{
         \subfloat[Case I]{\includegraphics[height=2.5in,width=1.2in]{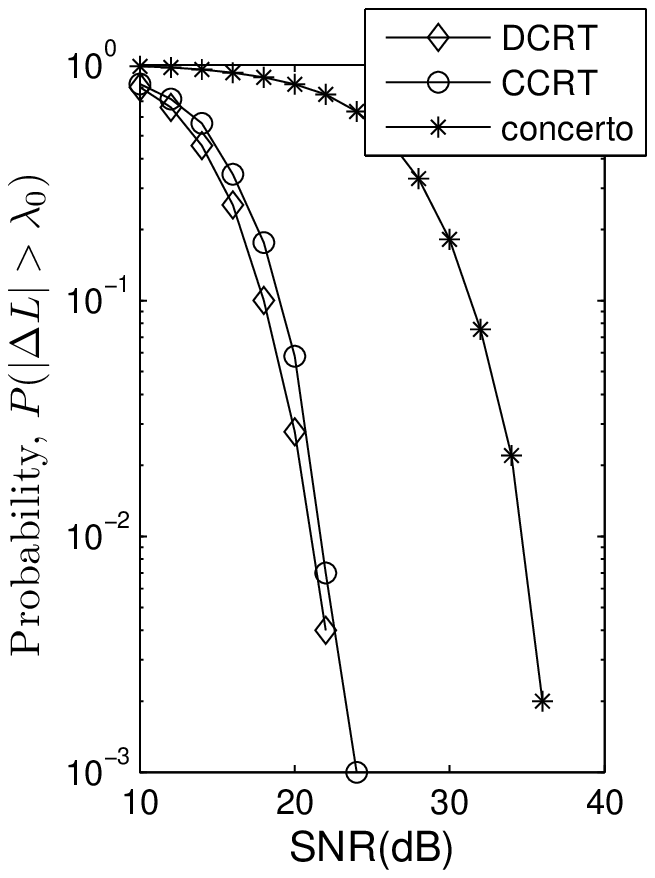}%
            \label{fig_pf_N4}}\hspace{-2mm}
         \subfloat[Case II]{\includegraphics[height=2.5in,width=1.2in]{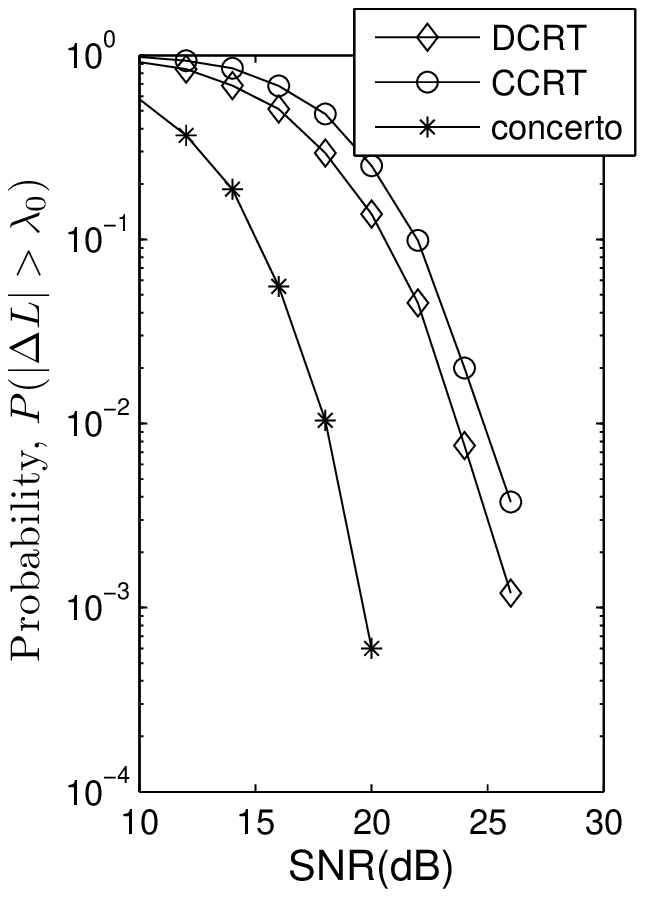}%
            \label{fig_pf_N6}}
          \hspace{-2mm}
         \subfloat[Case III]{\includegraphics[height=2.5in,width=1.2in]{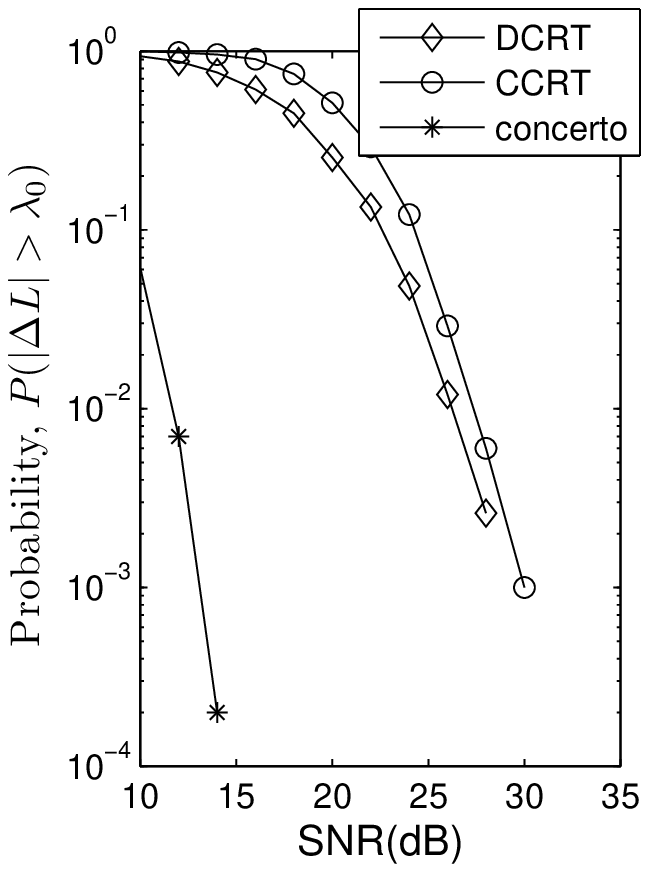}%
            \label{fig_pf_N8}}
         }
     \vfill
     \vspace{-0pt}
     \caption{Probability of the absolute error larger than the wavelength $\lambda_0$ with varying N. (a) $N=4$. (b) $N=6$. (c) $N=8$.}
     \label{fig_comp_ccrt_concert}
    \end{figure}

   In Fig.\ref{fig_comp_ccrt_concert}, we compare the reliability ($P(|\Delta L|>\lambda_0)$) of three closed-form method (the \textit{concerto} and two CRT-based estimator, i.e. DCRT and the closed-form CRT (CCRT) \cite{Wang:10}), in case of small number of frequencies, i.e. $N=4$, $N=6$ and $N=8$.
   Under the same bandwidth and frequency number, two set of frequency patterns (or wavelength patterns) are used to meet the requirement of each estimator.
   The proposed frequency pattern is used with $K=10000\mathrm{\,m}$ for \textit{concerto}.
   As a result, the wavelengths used for \textit{concerto} are ${\lambda}=\{ 1.1,1.1001,1.1075,1.9 \}$, ${\lambda}=\{ 1.1,1.1001,1.1011,1.1092,1.1849,2.9 \}$, and ${\lambda}=\{ 1.1,1.1001,1.1005,1.1023,1.1098,1.1433,1.3144,3.7 \}$ for $N=4$, $N=6$ and $N=8$.
   Correspondingly, the wavelengths patterns used for both DCRT and CCRT are
    ${\lambda}=\{ 1.1,1.3,1.7,1.9 \}$, $\mathbf{\lambda}=\{ 1.1,1.3,1.7,1.9,2.3,2.9 \}$,
   $\mathbf{\lambda}=\{ 1.1,1.3,1.7,1.9,2.3,2.9,3.1,3.7 \}$.
   The reliability of \textit{concerto} is dramatically improved with frequency number $N$.
   On the contrary, the reliability of both DCRT and CCRT are reduced rather than improved when frequency number is increased, with DCRT slightly outperforming CCRT.
   Therefore, DCRT and CCRT are more suitable for the case that the useable frequencies are very limited while \textit{concerto} benefits from an increase of frequency number.
   Moreover, \textit{concerto} is superior to the others even for relatively small number of frequencies, i.e. $N\geq6$.

     \begin{figure}[t]   
         \centering{\includegraphics[width=85mm,height=65mm]{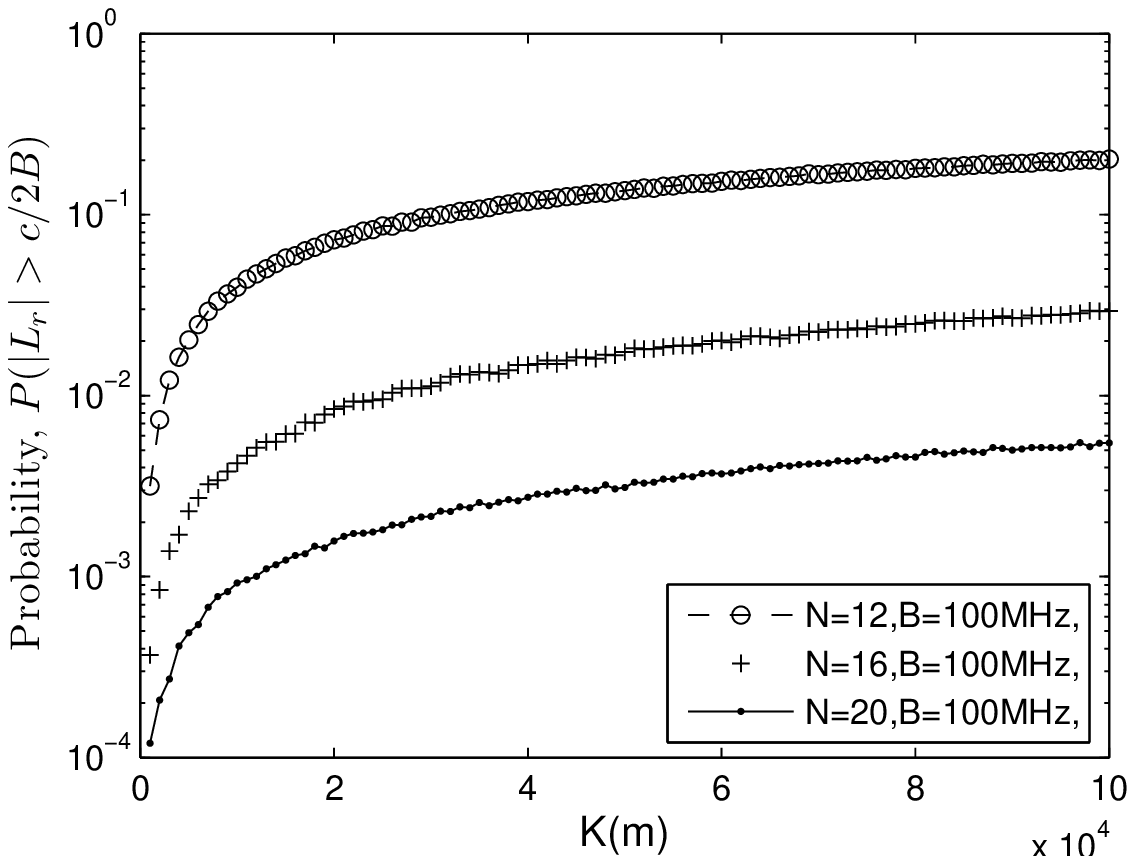}}
         \caption{ Probability $P(|L_r|>c/2B)$ versus $K$ under different $N$ with $\textrm{SNR}=5\textrm{dB}$.
           }
      \label{fig_K_N_SNR5}
    \end{figure}

   The assumption that the estimation error $L_r$ of  the first stage of \textit{concerto} satisfies $|L_r|<c/2B$, plays a great role in \textit{concerto}. To validate this hypothesis,
   Fig.\ref{fig_K_N_SNR5} shows the impact of measurement range $K$ on the probability $P(|L_r|>c/2B)$ for relatively low $\textrm{SNR}=5\textrm{dB}$. With $N=16$, the probability is observed to remain small even for $K$ as large as $100 \mathrm{\,km}$.
   In other words, with mild assumptions on $\textrm{SNR}$ and $N$, the measurement range of \textit{concerto} may be greatly extended by increasing $K$ while maintaining a very low probability $P(|L_r|>c/2B)$.



\begin{figure}[t]   
         \centering{\includegraphics[width=85mm,height=65mm]{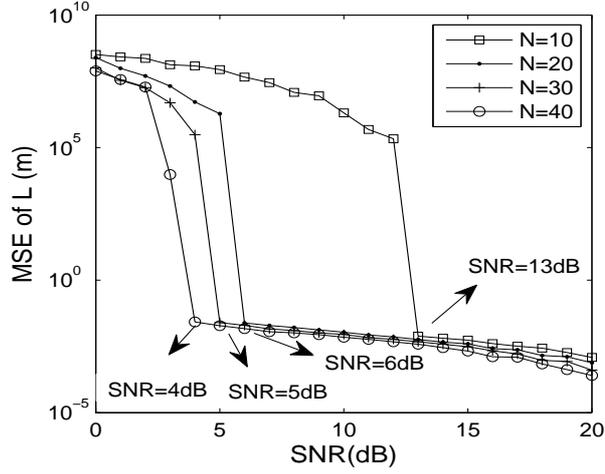}}
         \caption{SNR threshold of \textit{concerto} versus $N$ using the proposed frequency pattern with $K=100 \textrm{\,km}$ and $B=100\textrm{\,MHz}$.
           }
      \label{fig_threshold}
    \end{figure}

   In Fig.\ref{fig_threshold}, with a fixed $K=100 \mathrm{\,km}$, we investigate the impact of frequency number $N$ on SNR threshold of \textit{concerto}. If SNR is below the threshold, the estimation accuracy will deteriorate significantly and the result becomes completely useless
   \footnote{For more information on the threshold effect, see \cite{Athley:05}}.
   Therefore, the SNR threshold should be as low as possible. A remarkable improvement in SNR threshold is observed when frequency number is increased from $N=10$ to $N=20$. For $N$ larger than $20$, the improvement in threshold grows slowly. This result reveals that the frequency number required for \textit{concerto} to work is not too much, even for a quite large measurement range.

     \textit{Remark 3}: We observe that \textit{concerto} outperforms EF and DCRT in estimation accuracy, reliability and computation complexity \textit{simultaneously}. Meanwhile, an extremely large UMR can be guaranteed.

     Moreover, the signal model of INSAR \cite{Yuan:13} is the same as the one in this letter and the phase noise follows an hypergeometric distribution which can be approximated as a Gaussian one \cite{Lahiri:07}.
     Therefore, the proposed method can also be applied to phase unwrapping in INSAR besides its application in geodesy \cite{Kusy:06,Wang:11,Li:13}.

\section{Conclusion and future work}

   Combination of a modified BW method and two optimal estimations for MFPU is presented in the letter.
   The key idea behind \textit{concerto} is to exploit the modified BW for fast initial estimate and two optimal
   estimations for recovering the lost information caused by BW, thus maintaining the merit of both.
   As a result, \textit{concerto}  is highly attractive for its accurate and reliable distance estimation as well as extremely low complexity.

\section{Appendix}
\appendix

\section{derivation of (\ref{equa:L_wpa})}
\label{append:optimal_estimation}

     We rewrite (\ref{equa:cost_max}) as
     \begin{align}
        &\underset{\hat{L}_r}{\max} \left(\sum_{i=0}^{N-1} \hspace{-2pt} e^{ j \left( \frac{2\pi f_i \hat{L}_r}{c}-\tilde{\phi}_{i} \right) } \right)
        \left(\sum_{k=0}^{N-1} \hspace{-2pt} e^{-j \left( \frac{2\pi f_k \hat{L}_r}{c}-\tilde{\phi}_{k} \right)  }    \right)\notag\\
        \label{equa:cost_Re}
        =&\underset{\hat{L}_r}{\max} \,\mathbf{Re} \Bigg\{ \sum_{i=0}^{N-2} \sum_{k=i+1}^{N-1}
        \hspace{-3pt} e^{ j\Big( \frac{ 2\pi \left(f_i-f_k \right)\hat{L}_r }{c}
         - \big(\tilde{\phi}_{i}- \tilde{\phi}_{k} \big) \Big) } \Bigg\}
      \end{align}
     Define
    \begin{align}
        &\phi_{i,k} (\hat{L}_r)=\frac{2\pi \left(f_i -f_k \right)\hat{L}_r}{c} -\left[ \tilde{\phi}_{i}- \tilde{\phi}_{k} \right]_{2\pi}\notag\\
        &=\frac{2\pi \left(f_i -f_k \right)\hat{L}_r}{c} -\Delta\tilde{\phi}_{i,k}\notag\\
        &=\frac{2\pi \left(f_i -f_k \right)\hat{L}_r}{c} -\left[\frac{2\pi \left(f_i -f_k \right){L}_r}{c}
        +\theta_e(i)-\theta_e(k)\right]_{2\pi} \notag\\
        \label{equa:de_mod}
        &=\frac{2\pi \left(f_i -f_k \right)\hat{L}_r}{c} -\left(\frac{2\pi \left(f_i -f_k \right){L}_r}{c}
        +\theta_e(i)-\theta_e(k)\right)
    \end{align}
    where
    $\Delta\tilde{\phi}_{i,k}=\left[\tilde{\phi}_{i}-\tilde{\phi}_{k}\right]_{2\pi}$.
    The last equality holds since $\left|L_r \right|< \frac{c}{2B} < \left| \frac{c}{2(f_i-f_k)} \right|$.
    According to (\ref{equa:de_mod}), the "good" $\hat{L}_r$ must satisfy $|\phi_{i,k} (\hat{L}_r)| \ll 1$.
    Using a second-order approximation of Taylor series $f(x)=e^{jx}$ at $x_0=0$,
    \begin{align*}
       f(x) &\approx f(x_0)+\frac{f'(x_0)}{1!}(x-x_0)+\frac{f''(x_0)}{2!}(x-x_0)^2 \\
       e^{jx}&\approx 1+jx-\frac{1}{2}x^2
    \end{align*}
    we have
    \begin{align}
      \label{equa:Taylor}
      \mathbf{Re}  \left\{   \sum_{i=0}^{N-2} \sum_{k=i+1}^{N-1} \hspace{-3pt}e^{ j\phi_{i,k} (\hat{L}_r) }   \right\}
      \hspace{-2pt}\approx\hspace{-2pt} \sum_{i=0}^{N-2}\sum_{k=i+1}^{N-1}\hspace{-3pt}\left(1\hspace{-2pt}-\hspace{-2pt}\frac{1}{2}\left( \phi_{i,k} (\hat{L}_r)\right)^2  \right)
    \end{align}
    Thus, (\ref{equa:cost_Re}) is equivalent to
    \begin{gather}
        \label{equa:cost_min}
        \hspace{-10pt}\underset{\hat{L}_r}{\min} \sum_{i=0}^{N-2} \sum_{k=i+1}^{N-1}  \left( \phi_{i,k} (\hat{L}_r) \right)^2
    \end{gather}
    The problem can be expressed as {\cite{Wu:95}}
    \begin{gather}
       \hspace{-10pt}\underset{ \hat{L}_r}{\min} \sum_{i=0}^{N-2} \sum_{k=i+1}^{N-1}  \left(\phi_{i,k} (\hat{L}_r) \right)^2
                           = \underset{\hat{L}_r}{\min}  \mathbf{\Phi}^T(\hat{L}_r)\mathbf{R}^{-1} \mathbf{\Phi}(\hat{L}_r)
    \end{gather}
    where
        $\mathbf{\Phi}(\hat{L}_r)=\left[\phi_{0,1} (\hat{L}_r) ,\phi_{0,2} (\hat{L}_r) ,\cdots
        \phi_{0,N-1} (\hat{L}_r) \right]^T$,
    and $\mathbf{R}^{-1}$ is defined in (\ref{equa:R_u}).

    The optimal solution $L_r^*$ obey
    \begin{align}
       \frac{ \partial \left(  \mathbf{\Phi}^T(\hat{L}_r)\mathbf{R}^{-1} \mathbf{\Phi}(\hat{L}_r)  \right)  }  {\partial (\hat{L}_r)}\Big|_{\hat{L}_r=L_r^*} &= 0 \notag\\
       \left( \partial \left[\mathbf{\Phi}(\hat{L}_r) \right]  /{\partial (\hat{L}_r)}  \right)^T  \mathbf{R}^{-1}  \mathbf{\Phi}(L_r^*)&= 0
    \end{align}
   Since $\partial \left[\phi_{0,i} (\hat{L}_r) \right] / \partial(\hat{L}_r)=\nobreak \frac{2\pi}{c}\left(f_0-f_i \right) =\nobreak \frac{2\pi}{c}\sum_{k=1}^{i} {\Delta f_k}$, where $\Delta f_k=f_{k-1}-f_k $, and
    \begin{align}
        \frac{\partial \left[\mathbf{\Phi}(\hat{L}_r ) \right]}{{\partial (\hat{L}_r)}}
        \hspace{-3pt}=\hspace{-3pt} \frac{2\pi}{c}\left[\Delta f_1,\Delta f_1+\Delta f_2,\cdots\sum_{k=1}^{N-1} {\Delta f_k } \right]^T
        \hspace{-3pt}=\hspace{-3pt} \frac{2\pi}{c}\mathbf{\Gamma }\Delta \mathbf{f}
    \end{align}
   It follows that
    \begin{align}
        \Delta \mathbf{f}^T \mathbf{\Gamma}^T \mathbf{R}^{-1} \mathbf{\Phi}\left(L_r^*\right) &= 0 \notag\\
        \Delta \mathbf{f}^T \mathbf{\Gamma }^T \mathbf{R}^{-1} \left( \frac{2\pi L_r^*}{c}
                 \mathbf{\Gamma} \Delta \mathbf{f} - \mathbf{\Gamma}\Delta\tilde{\mathbf{\Phi}} \right)&= 0
    \end{align}
   Then
    \begin{align}
         L_r^* &= \frac{c}{2\pi} \frac{\Delta \mathbf{f}^T \mathbf{\Gamma}^T \mathbf{R}^{-1} \mathbf{\Gamma} \Delta\tilde{\mathbf{\Phi}} }
                  { \Delta \mathbf{f}^T \mathbf{\Gamma}^T \mathbf{R}^{-1} \mathbf{\Gamma}\Delta \mathbf{f}  }
    \end{align}

\section{Proof of (\ref{equa:CRB})}   
\label{append:CRB}

     It is well known that the phase noise in (\ref{equa:basic_noise}) follows the wrapped normal distribution due to
     modulo $2\pi$ operation\cite{Cheng:11}. So the estimation is usually not unbias and the CRB does not exist. But the
     noise can be approximated as normal distribution under the assumption of high SNR.

     Consider the signal $y(k)= Ae^{j2\pi L / \lambda _k }+ n(k)$, where $L$ is the parameter to be estimated and  $n(k)$ is the complex Gaussian noise with zero-mean and variance $\mathbf{E}[ n^2(k)]=\sigma_n^2$ and the signal-to-noise ratio (SNR) is defined as $\textrm{SNR} =A^2/\sigma_n^2$. The signal can be expressed as
     \vspace{-0pt}
     \begin{align}
        \hspace{-8pt} y(k) &= Ae^{j2\pi \frac{L}{\lambda _k}   }  + n(k) \notag\\
                           &= Ae^{j2\pi \frac{L}{\lambda _k} } (1 + n(k) e^{-j2\pi \frac{L}{\lambda _k} }/ {A} )
     \end{align}
      \vspace{-0pt}
     Let $n'(k) = n(k)e^{-j2\pi L / \lambda_k }/A = n'_R (k) + n'_I (k)j$, then $\mathbf{E}[n'^2 (k) ]=\sigma_n^2/A^2$ and $\mathbf{E}[{n'_I}^2(k)] = {\sigma_n^2/(2A^2)}$. Therefore
     \begin{align}
        y(k)&=Ae^{j2\pi L /{\lambda _k }} (1+n'_R(k)+n'_I(k)j) \notag\\
            &=A \sqrt { (1+n'_R(k))^2+{n'_I}^2(k) } e^{(j2\pi L /{\lambda _k }+\theta_e (k) )}
     \end{align}
      where $\theta_e (k)$ is phase noise. At high SNR, the following approximation holds
     \begin{gather}
     \theta_e(k) \approx tan(\theta_e (k))= \frac{ n'_I (k)} {1 + n'_R (k)} \approx n'_I (k)\\
     \mathbf{E} \left[\theta_e^2 (k) \right] = \sigma^2  \approx {\sigma_n^2 /(2A^2)}
     \end{gather}

     Define $\mathbf{y} = [y(1),y(2),\cdots y(N) ]$ and $\mathbf{\varphi} =[\varphi(1),\varphi(2),\cdots \varphi(N) ]$, $\varphi (k) =[{2\pi L/\lambda_k+\theta _e(k)}]_{2\pi}$. It is clear that the original problem of estimating $L$ using $\mathbf{\varphi}$, corrupted by noise of variance ${\sigma_n^2 /(2A^2)}$, is \textit{equivalent to} estimating it from noisy signal $\mathbf{y}$ with noise variance $\sigma_n^2$. The Cramer-Rao bound of the equivalent problem is easily obtained as follows.

     Let  $a_k =\mathbf{Re} \{y(k)\}$, $b_k =\mathbf{Im} \{y(k)\}$. The probability distribution function is\cite{Rife:74}
     \begin{align}
        f(\mathbf{y},L)&\hspace{-2pt}=\hspace{-4pt}\left(\frac{1}{\sqrt {\pi \sigma_n^2} }\right)^{2N}
                         \hspace{-4pt}\exp \hspace{-3pt}\Bigg[\hspace{-4pt}-\hspace{-2pt} \frac{1}{\sigma _n^2}\sum_{k=1}^N  \bigg( \big(a_k \hspace{-3pt}- \hspace{-3pt}A\cos ( 2\pi L /\lambda _k) \big)^2  \notag\\
                         &\hspace{70pt}+ \big(b_k- A\sin ( 2\pi L /\lambda _k) \big)^2  \bigg) \Bigg]
     \end{align}
     The entry of Fisher information matrix and the Cramer-Rao bound (CRB) for $L$ estimation in MFPU are given by
     \begin{align}
        \hspace{-8pt}[\mathbf{F}]_{L,L}\hspace{-1pt}&=\hspace{-1pt}\mathbf{E}\left[ \frac{\partial \log f(\mathbf{y},L)}{\partial L}
        \frac{\partial \log f(\mathbf{y},L)}{\partial L} \right]\hspace{-1pt}  \notag \\
                                                 &=\hspace{-1pt}\frac{8A^2\pi^2}{ \sigma_n^2 } \sum_{k=1}^N {\lambda_k^{-2} }\\
        \mathrm{CRB}(L)&= ([\mathbf{F}]_{L,L})^{-1}
                       = \frac{\sigma ^2}{4\pi^2}\left(\sum_{k=1}^N {\lambda_k^{-2}} \right)^{-1}
     \end{align}




\end{document}